\documentclass[letter]{aa}

\usepackage{lscape}
\usepackage{graphics,graphicx}
\usepackage{natbib}
\usepackage{longtable}
\usepackage{supertabular}
\bibpunct{(}{)}{;}{a}{}{,}

\usepackage{txfonts}
\begin{document}

   \title{Discovery of X-ray flaring activity in the Arches cluster}
   \titlerunning{X-ray flaring activity in the Arches cluster}
   \authorrunning{R. Capelli}

   \author{R. Capelli\inst{1}, R.S. Warwick\inst{2}, N. Cappelluti\inst{1},  S. Gillessen\inst{1}, P. Predehl\inst{1}, D. Porquet\inst{3} and S. Czesla\inst{4}}

   \offprints{R. Capelli,\\ e-mail: capelli@mpe.mpg.de}

   \institute{Max-Planck-Institut f\"ur Extraterrestrische Physik, 
              Giessenbachstrasse 1, 85748 Garching, Germany
\and
	      Department of Physics and Astronomy, University of 
	      Leicester, Leicester LE1 7RH, UK
\and
	      Observatoire Astronomique de Strasbourg, Universit{\'e} de Strasbourg, CNRS, UMR 7550, 11 rue de l'Universit{\'e}, F-67000 Strasbourg, France
\and
	      Hamburger Sternwarte, Universit\"at Hamburg, Gojenbergsweg 112, 21029 Hamburg, Germany
}

\date{Received ...; ...}

\abstract{We  present a study of the Arches cluster based on  
XMM-Newton  observations  performed over the past 8 years. 
Unexpectedly, we find that the X-ray emission associated with the cluster
experienced a marked brightening in March/April  2007.}
{We investigate the origin of both the X-ray continuum emission
emanating from the star cluster and the flare.}   
{To study the  time variability of  the total X-ray  flux, we
stacked the PN and MOS data of  observations performed within a time interval
of a few days leading to the detection of the flaring episode.  We then constructed
two  spectral datasets,  one corresponding to the flare interval (March/April 2007) 
and another to the normal quiescent state of the source.}   
{The X-ray light curve of the  Arches cluster  shows, with high
significance (8.6$\sigma$), a  70\% increase in the X-ray  emission 
in the March/April 2007 timeframe
followed by a decline over the following year to the pre-flare level; 
the short-term duration of the flare is constrained to be longer 
than four days. The temperature  and 
the line-of-sight column density inferred from the flare  spectrum 
do not  differ from  those measured in  the normal activity state 
of the cluster, suggesting that the flux enhancement is thermal in
origin.}   {We  attribute the  X-ray  variability to  in situ stellar
activity: early-type stars may be responsible for the flare
via  wind collisions,  whereas late-type  stars may  contribute by means of
magnetic reconnection. These two possibilities are discussed. }

% 5 {} token are mandatory


   \maketitle

% =============================================================================
\section{Introduction}
Three  massive  star  clusters  are  located at  the  Galactic  center
(hereafter GC) region:  the central (IRS 16) star  cluster, the Arches
cluster (hereafter  AC), and    the     Quintuplet    cluster
\citep[e.g.,][]{1996ARA&A..34..645M}.  Whereas   all  three  are  very
bright   in  the  infrared   band  \citep[e.g.][]{1975ApJ...200L..71B,
1995AJ....109.1676N},  their  X-ray  properties differ  markedly.  
Both IRS16 and the Quintuplet cluster are not particularly bright in X-rays 
with L$_{X}$$\sim$10$^{33}$erg/s \citep[][]{2003ApJ...591..891B,2004ApJ...611..858L}.
In constrast, the AC has a relatively high X-ray luminosity 
(L$_X \sim 10^{34}$erg/s)\citep[e.g.][]{2002ApJ...570..665Y}.
 Although  the  total
mass  of each  of  these clusters  is  similar at about 10$^{4}$  M$_{\odot}$
\citep[][]{2004ASPC..322...49F},   the  size   of  the   clusters  are
different,  with the  AC  being  the most  compact.  The X-ray  diffuse
emission from  the AC has been investigated in a number of published 
papers over the last decade  
\citep[i.e. ][]{2002ApJ...570..665Y,  2006MNRAS.371...38W,
2007PASJ...59S.229T}.  All  these  studies confirm  the
co-existence  in   the  cluster   of  both  thermal   and  non-thermal
radiation. The  former is thought  to be due largely to  
multiple interactions between  the strong  winds from  massive  stars, 
with the high stellar  concentration in the AC providing an explanation
of why its X-ray  luminosity is enhanced 
\citep[e.g. ][]{2000ApJ...536..896C,
2001ApJ...559L..33R}.  Irrefutably,   the  AC, which is the  densest 
cluster in  the Galaxy  with a  core density  of 10$^{5}$
M$_{\odot}$/pc$^{3}$,  provides   an  excellent opportunity 
to study stellar processes, within the environment of a massive 
star cluster, that lead to particle acceleration and X-ray emission.
The study  of these
massive star clusters is of  particular interest in the context of the
chemical enrichment of the GC region and their overall contribution to 
the high energy activity that characterises the innermost zone of the
Galaxy.

\section{Observations and data reduction}

We  selected archival XMM-Newton observations of  the Galactic  center 
region targeted at  Sgr A*. We reprocessed  data from both the PN and 
MOS cameras (\citet{2001A&A...365L..18S},
\citet{2001A&A...365L..27T}) with  the tasks EPPROC and  EMPROC in the
Science Analysis Software version 9.0. The main properties of the XMM-Newton
observations  employed  in the  present  work  are  reported in  Table
\ref{log_table}.  We selected  good  time intervals  (GTI) corresponding
to periods of relatively low  internal background. For this purpose,
we compiled the 10--12  keV lightcurve of  the full field of view and
selected  the  time intervals  for  which   the  count  rate  was  lower   
than  a  certain threshold.  
Since the  selected  observations were performed  in
different  conditions  of  internal/particle  background  and  orbital
phase, we investigated all  the lightcurves independently and selected
an appropriate  threshold   count    rate   for each.
The  thresholds used  in  screening  the  data are reported in Table 
\ref{log_table}; in practice, these thresholds were selected so
as to exclude all the peaks in the 10--12 keV full-field lightcurve. 
Throughout our
analysis,   we   have   only   selected  single   and   double   events
(PATTERN$\le$4) for  the PN and up to  quadruple events for  the
MOS1 and MOS2 (PATTERN$\le$12) cameras. Similarly, only 
events tagged as real X-rays (FLAG==0) were utilised.

\onltab{1}
{
\begin{table*}
\caption{The selected datasets identified by both the
OBSID and the observation date. For each instrument, we report the 
threshold in the 10-12 keV lightcurve (counts/s) used for GTI selection, 
the total GTI exposure (ks) and the nominal observation duration (ks).} 
\label{log_table}
\centering
\begin{tabular}{c|c|ccc}
\hline
\hline
OBSID & Obs Date & PN & MOS1 & MOS2 \\
 & yyyy-mm-dd &  cut/GTI/exp & cut/GTI/exp & /GTI/exp \\
\hline
0111350101 & 2002-02-26 & 0.8/38.590/40.030 & 0.5/42.262/52.105 & 0.5/41.700/52.120 \\
0202670501 & 2004-03-28 & 2.0/13.320/101.170 & 1.0/33.070/107.784 & 1.0/30.049/108.572 \\
0202670601 & 2004-03-30 & 2.0/25.680/112.204 & 1.0/32.841/120.863 & 1.0/35.390/122.521 \\
0202670701 & 2004-08-31 & 1.0/59.400/127.470 & 0.5/80.640/132.469 & 0.5/84.180/132.502 \\
0202670801 & 2004-09-02 & 1.0/69.360/130.951 & 0.5/94.774/132.997 & 0.5/98.757/133.036 \\
0402430301 & 2007-04-01 & 1.5/61.465/101.319 & 0.8/61.002/93.947 & 0.8/62.987/94.022 \\
0402430401 & 2007-04-03 & 1.5/48.862/93.594 & 0.8/40.372/97.566 & 0.8/41.317/96.461 \\
0402430701 & 2007-03-30 & 1.5/32.337/32.338 & 0.8/26.720/33.912 & 0.8/27.685/33.917 \\
0505670101 & 2008-03-23 & 1.25/74.216/96.601 & 0.5/73.662/97.787 & 0.5/74.027/97.787 \\
0554750401 & 2009-04-01 & 1.0/30.114/38.034 & 0.5/32.567/39.614 & 0.5/33.802/39.619  \\
0554750501 & 2009-04-03 & 1.0/36.374/42.434 & 0.5/41.376/44.016 & 0.5/41.318/44.018 \\
0554750601 & 2009-04-05 & 1.0/28.697/32.837 & 0.5/37.076/38.816 & 0.5/36.840/38.818\\
\hline
\end{tabular}
\end{table*}
}

It was evident  that the source was seen in an  anomalous state in the
three 2007 observations (separated by four days).  For this reason, we
considered the March/April 2007 observations (i.e.,
the  flare dataset)  separately from the other  observations.  The region
considered  for the  accumulation of  the count  rate spectrum  of the
source is  an ellipse centered  at R.A.=17:45:50.358 DEC=-28:49:20.08,
with axes of 0.32'x0.30'. We chose an elliptical region in order to enclose most of the 
X-ray diffuse emission in the AC \citep[see contours in Fig.2 in][]{2002ApJ...570..665Y}.
Similarly a circular region of radius 0.4',
centered on  R.A.=17:45:44.081 DEC=-28:49:25.72 was  used to determine
the  background. Spectra corresponding  to multiple  observations were
produced by  stacking the  data, using the  tools MATHPHA,  ADDRMF and
ADDARF. The  channels in  each spectrum were  grouped with  the GRPPHA
tool in order  to have a minimum of 20  counts/bin, validating the use
of the $\chi^{2}$ statistic.

\section{Results}
\subsection{The X-ray lightcurve}

To search for evidence of temporal variability in the
X-ray emission from the AC, we constructed a ``long-term'' X-ray
lightcurve based on six observation epochs, which in the majority of cases 
encompassed more than one individual observation. 
These were February 2002 (1 obs), March 2004 (2 obs), 
August/September 2004 (2 obs), March/April 2007 (3 obs), March 2008 (1 obs),
and April 2009 (3 obs) - see Table \ref{log_table}.

\begin{figure}[ht]
\begin{center}
\includegraphics[width=0.35\textwidth]{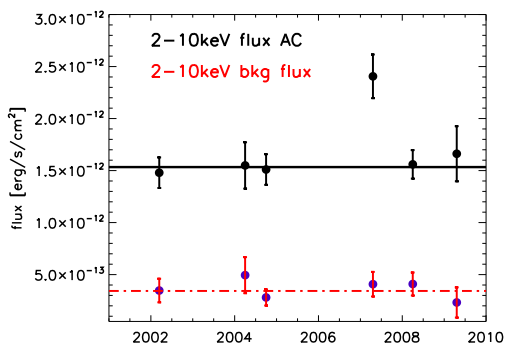}
\end{center}
\vspace{-0.5cm}
\caption{The 2--10 keV X-ray lightcurve of the AC extending 
over an eight year period. 
The horizontal line is the weighted mean excluding the March/April 2007
flare. The red points show the lightcurve for the background, where the weighted mean is shown with a dashed red line.}
\label{lc_AC}
\end{figure}

The resulting 2--10 keV lightcurve is  shown in  Fig.\ref{lc_AC}.
Here the flux values are derived from model fits to the
stacked spectra for each epoch using the average spectrum for the
normal state (see below); more specifically, the points
are the weighted mean of  the PN and MOS values,  corrected
for absorption. A high state or flaring episode is evident in 
the 2007 measurements. As a check, we also considered the net source count rates 
measured in the individual cameras, which confirmed that this flaring is very likely a feature of the lightcurves. To  quantify  the  statistical
significance of  this flare,  we compared the  weighted mean  of the
other five measurements in Fig.\ref{lc_AC}
of 1.53$\pm$0.07$\times$10$^{-12}$ erg/cm$^{2}$/s 
with  the   flux  measured   in  March/April   2007   of  2.4$\pm$0.2
$\times$10$^{-12}$  erg/cm$^{2}$/s;   the  variability is significant
at the level of  8.6$\sigma$.

%-----------------------------Figure Start------------------------------
%-----------------------------Figure End--------------------------------

To check whether the observed flare was produced by systematics in the background, we also built the lightcurve of the flux measured from the region used for the background (the red points in Fig.\ref{lc_AC}). For these data, the $\chi^{2}_{red}$ with respect to the weighted mean of the measurements (3.4$\pm$0.4$\times$10$^{-13}$erg/cm$^{2}$/s) is 0.7, fully consistent with a constant background.

\begin{figure}[ht]
\begin{center}
\includegraphics[width=0.35\textwidth]{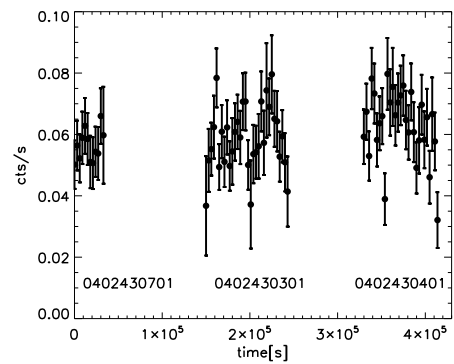}
\end{center}
\vspace{-0.5cm}
\caption{The lightcurve from the March-April 2007 observations. The energy range is 2--10 keV and 
the time binning is set to 3 ks.}
\label{single-lc}
\end{figure}

We have also studied the 2--10 keV lightcurve of the individual March-April 2007 observations to look 
for evidence of short-term variability during the episode of enhanced activity. The result of this investigation is shown in Fig.\ref{single-lc}. There is no clear evidence of a trend in this full lightcurve; a constant fit to the data gives a mean count rate of 0.059$\pm$0.001 cts/s, with a $\chi^{2}_{red}$=1.1. Although there are some low-level short-term features apparent in the individual observations, we conclude that these are most likely due to background-subtraction errors, since similar effects are apparent in the lightcurves of other sources in the
field of view. We conclude that the duration of the high state activity is longer than $\sim$4 days.

%-----------------------------Figure Start------------------------------
%-----------------------------Figure End--------------------------------

\vspace{-0.5cm}

\subsection{X-ray spectral analysis}

We first performed an analysis of the stacked PN and MOS X-ray 
spectra pertaining to the normal state of the AC. The spectral model
comprised a  collisionally ionized plasma  
\citep[APEC,][]{2001ApJ...556L..91S} plus a non-thermal  hard tail  
modeled by a power-law continuum with a photon index $\Gamma$.  In the present investigation,
we found it advantageous to fix $\Gamma$ to a value of 1, thereby avoiding potential 
biases in the spectral modeling of multiple datasets with a different balance
between the thermal and non-thermal contributions. This approach is consistent with 
earlier studies that generally found  $\Gamma$ to be in  the range 1-1.5
\citep[e.g.,][]{2006MNRAS.371...38W,  2007PASJ...59S.229T} and consistent
with the power-law slope characterising bright 6.4 keV fluorescent line-emitting 
regions in the vicinity of the AC (which will be the subject of a later study).
In the analysis, we also fixed the metallicity of the thermal plasma
at 2 Z$_{\odot}$  \citep[e.g.,][]{2006MNRAS.371...38W}, 
since the spectral fitting of both the stacked and the 
single datasets did not usefully constrain this parameter. In this context, 
we note that there is strong evidence that the diffuse plasma in the GC region has 
a super-solar metallicity \citep[e.g.,][]{2002A&A...382.1052T,2007PASJ...59S.245K}.
Two gaussian  lines   were included to  account  for the neutral (or near-neutral)  Fe
K$_{\alpha}$ and  Fe K$_{\beta}$ lines,  respectively at 6.4  and 7.05
keV, the K$_{\beta}$/K$_{\alpha}$ flux ratio being tied to 0.11.  
Finally  all the components were subject to interstellar
absorption  \citep[WABS model  in  XSPEC,][]{1983ApJ...270..119M}. 

\begin{figure*}[!Ht]
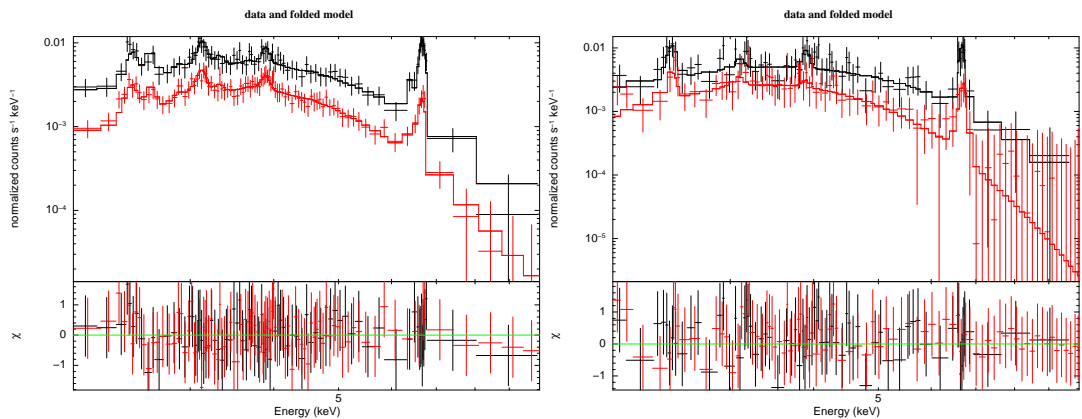

\begin{center}
% un-comment the following line to include your fig1a.ps postscript file
\includegraphics[width=0.3\textwidth,angle=-90]{fig3a.ps}
\includegraphics[width=0.3\textwidth,angle=-90]{fig3b.ps}
\end{center}
\vspace{-0.5cm}
\caption{The 2--10 keV X-ray spectra of the AC as measured in the
PN (black) and MOS (red) cameras. {\it Left panel:} The spectrum measured
in the normal state.  {\it Right panel:} The net flare spectrum measured during
the flaring episode.  In both cases, the best fitting spectral model 
and the corresponding residuals are also shown.}
\label{AC_spec}
\end{figure*}

The results  of the joint fitting of the PN and MOS spectra
for the normal state of the source are reported  in the second column 
of  Table \ref{source_fitting} and the corresponding spectra are shown in
Fig.\ref{AC_spec} (left panel). The
best fit temperature of the cluster was found to be kT $=$1.7$\pm$0.1
keV,  in  good  agreement  with  what  previously  
measured with  Suzaku
\citep[][]{2007PASJ...59S.229T}               and              Chandra
\citep[][]{2006MNRAS.371...38W}.  Moreover, the  resulting  total 2--10
keV     flux    is     5.3$\pm$0.5$\times$10$^{-13}$    erg/cm$^{2}$/s
(1.5$\pm$0.1$\times$10$^{-12}$  erg/cm$^{2}$/s  absorption corrected),
translating into a luminosity of 3.8$\pm$0.4$\times$10$^{33}$erg/s.
To quantify the respective contributions  of the thermal  and non-thermal
emission to  the spectrum,  we measured the  fluxes associated
with the  different continuum  components in the  best fit  model.
For the 2--10 keV band, the split was 85\% contributed by the thermal 
emission and 15\% by the hard non-thermal tail.
Both the thermal and non-thermal components are very likely to be
manifestations of in situ activity within the  AC. More specifically  
shocks produced  by strong  colliding stellar winds can both heat the 
plasma and supply high  energy particles.
\citet[][]{2004ApJ...611..858L}    proposed   that   energetic
electrons within the  cluster might be able to  upscatter IR radiation
from  stars via  Inverse Compton  Effect; in  this case,  the power-law
component  should  contribute  about a sixth of  the  total  flux,
in excellent agreement with the measurements.

\vspace{-0.15cm}

%*****************************************************************************
\begin{table}[ht]
\caption{Results of the spectral fitting for the normal state and the net flare spectrum of the AC. 
The values are the weighted mean of the PN and MOS measurements. The fluxes (2--10 keV) are in photons/cm$^{2}$/s. }
\label{source_fitting}
\centering
\begin{tabular}{c|cc}
\hline
\hline
 & NORMAL & FLARE \\
\hline
N$_{H}$(10$^{22}$cm$^{-2}$) & 10.4$\pm$0.9 & 10.2$\pm$2.1 \\
kT $_{APEC}$(keV) & 1.7$\pm$0.1 & 1.8$\pm$0.3 \\
flux$_{APEC}$(10$^{-5}$) & 5.4$\pm$0.4 & 5.2$\pm$0.9 \\
$\Gamma_{POW}$ & 1.0 & 1.0 \\
flux$_{POW}$(10$^{-5}$) & 1.3$\pm$0.1 & - \\
f$_{Fe_{I}K_{\alpha}}$(10$^{-6}$) & 1.4$\pm$0.7  & -    \\
flux(10$^{-5}$) & 7.0$\pm$0.5 & 5.2$\pm$0.9 \\
$\chi^{2}$/dof & 234/1144 & 246.5/852 \\
\hline
\end{tabular}
\end{table}
%**************************************************************************

%-----------------------------Figure Start------------------------------
%-----------------------------Figure End--------------------------------

Given  the discovery of significant variability of the X-ray flux
from  the AC, the next step was to investigate the spectrum
of the enhancement.  For this we used  the stacked 2007  dataset as the
source event file, with the background taken to be the source
spectrum of the  AC during  its normal state. 
The PN and MOS spectra corresponding to the flare
(net flare spectra) are presented  in the
Fig.\ref{AC_spec} (right panel), in comparison to
the normal  state  spectrum (left  panel).  
The  main difference  between  the spectra of the two states of the AC 
is in the 6.4 keV fluorescent line from neutral (or near-neutral) Fe; the flare spectrum 
does not show significant emission at this energy.  This  clearly  
indicates  that the 6.4 keV Fe
K$_{\alpha}$ line  has a different  origin to the  higher ionization
6.7 keV iron line.  This observational evidence points to a thermal
origin for the flare. This was confirmed when we attempted to fit
the flare spectra with the same  model as applied to the normal state
data. For the flare we found that the  nonthermal power-law 
and the low-ionization state  lines of Fe  (at 6.4 and  7.05 keV) were  no more
required. The flare  spectrum is well described by a thermal plasma with  
kT=1.8$\pm$0.3 keV with the total flux being
3.7$\pm$0.6$\times$10$^{-13}$  erg/cm$^{2}$/s; in  terms  of 2--10  keV
X-ray  flux, the  flare represents an increase over the normal state
activity of  about 70$\pm$13\%. 

We highlight that  the temperature and the N$_{H}$  values inferred
for the flare  spectra are in excellent agreement with the values
derived from the normal state  spectrum (see Fig.\ref{AC_spec}
and Table  \ref{source_fitting}). It would seem that the  variability in
the cluster is produced by an increase in  the intensity of the thermal
emission,  while the  other  physical  observables, namely the
temperature and N$_{H}$ remain  unchanged. The emission measure (EM) of
this X-ray activity (the flare component) has been estimated from the normalization of
the APEC component to be some 1.2$\times$10$^{57}$cm$^{-3}$.

\section{Discussion}
In  this paper we  have investigated the  X-ray properties of the AC
 and report the first detection of significant variability  in  
its  hot thermal  emission.
The bulk of this component is thought to arise from the
thermalization  of  the  strong   winds  expelled  by  massive  stars.
\citet[][]{2000ApJ...536..896C} calculated the expected temperature of
such a cluster wind to be

\[\mathrm{ kT_{cluster}=1.3 \left( \frac{v_{w}}{10^{3}km \cdot s^{-1}}\right)^{2}  keV }
\]

where v$_{w}$ is  the terminal velocity for the  stellar wind, which is assumed
to be the  same for all the stars within  the cluster. The temperature
we measured in the AC is  1.7$\pm$0.4 keV, which requires 
a terminal velocity  of 1100-1200 km/s, in good agreement with 
the actual wind velocities reported by
\citet[][]{1996ApJ...461..750C}.
In the normal state of the AC, the EM  of this
plasma is 1.4$\pm$0.3$\times$10$^{57}$cm$^{-3}$. If we model the AC as
a  sphere of  0.7 pc  radius and set  n$_{e}$$\sim$n$_{H}$,  the inferred
electron  density is 1-10 cm$^{-3}$.  This value again compares
very well to theoretical estimates obtained from
modeling  the  X-ray  emission  of  dense clusters  of  massive  stars
\citep[see   middle   panel   of   Fig.3   in][]{2000ApJ...536..896C}.
Using the superior spatial resolution of the  Chandra telescope,
\citet[][]{2002ApJ...570..665Y} identified three distinct spatial
components in the AC; components A1 and A2 were found to have strong thermal emission,
whereas A3 component was found to be extended  and more
elongated  towards the south-west  of the  cluster core.  Overall the diffuse
emission is  
characterized by  a  highly  significant  Fe  
K$_{\alpha}$ line at 6.4 keV corresponding to (neutral or  low
ionization gas). Because of the lower angular
resolution of  XMM-Newton, we  could not resolve  these three 
components and study them  separately but simply measure a composite
spectrum. 
On the same basis, \citet[][]{2007PASJ...59S.229T} have used
Suzaku data to study a 1.4 arcmin wide region centred on the cluster and
derive spectral parameters very similar to those reported here.  

The non-thermal radiation associated with the AC is  most likely attributed 
to the  presence of a sea of cosmic rays permeating the  whole region within the cluster and the
surroundings; \citet[][]{2000ApJ...543..733V}  showed that the diffuse
emission  from the Galactic  Ridge can  be explained  in terms  of Low
Energy Cosmic  Rays (LECRs) producing  both the continuum and  the low
ionization  lines  from  the  different  elements. However,
the  slope  of  the non-thermal  continuum  emission  found  by  these  
authors  is  about 1.3-1.4, somewhat steeper than the value we adopt.
This difference may reflect
the extreme conditions in the vicinity  of the AC itself. The collisions between
powerful stellar winds with terminal velocities at or in excess of 1200 km/s
can create  shocks which  accelerate  CRs  up to  keV energies, thereby
leading to a  flattening of the spectrum  in the high energy  domain via
either bremsstrahlung or inverse Compton emission.

Here we report an  unexpectedly high state of activity  in the  AC, 
although the XMM-Newton Off-axis PSF does not allow 
us to directly locate the flare emission within the spatial substructure observed in the
AC with Chandra \citep[][]{2002ApJ...570..665Y}.
The spectrum of  the AC in this flaring state  is well modeled as  
a hot plasma with  kT=1.8$\pm$0.3 keV.  This,  together with  the  presence in  the
spectrum of the high ionization lines from elements like S, Ar,
Ca  and Fe points to a  stellar origin  for the observed flux increase.
In the case of stellar coronal flares, the longest detected to date  is nine days.
For the AC we can set a lower  limit for the duration of the enhanced 
luminosity  of $\sim$4 days. This could fit with a coronal origin
for the flare. On the other hand, we notice that if the flare duration
is a few months, its origin is very unlikely to be coronal. Assuming a distance to the  GC of 8 kpc
\citep[][]{2009ApJ...692.1075G}, the estimated EM for the AC flare
is 1.2$\times$10$^{57}$  cm$^{-3}$, very significantly higher  than 
values which are typically measured  in  stellar   X-ray   flares   ($\sim$10$^{55}$
cm$^{-3}$) associated with late-type stars 
\citep[e.g.,][]{2004A&ARv..12...71G,2009A&ARv..17..309G}.
The X-ray flares from late type stars observed to date show strong correlation between 
the peak temperature and the EM. \citet[][]{2008ApJ...672..659A} found 
that the EM in a stellar flare is proportional to T$^{4.5}$; according 
to this relation, an EM of  1.2$\times$10$^{57}$  cm$^{-3}$ yields a 
temperature well above 10$^{8}$K (8.6 keV), well beyond the value measured 
in the AC flare (1.8 keV correspondent to $\sim$2$\times$10$^{7}$K). 
We therefore dismiss the late-type star hypothesis for origin of the flare.\\
An alternative scenario which could account for the detected variability
entails early-type stars, possibly in binary configurations. Either an orbital 
modulation of the system or a sudden eruption by one star might produce the observed
high state activity. In both cases the density and the velocities of
the colliding stellar winds might change, thus producing an enhancement 
in the X-ray flux.
The X-ray emission arising from equally strong colliding winds is inversely proportional 
to the binary separation (D$^{-1}$); for stellar parameters like the ones 
found in the AC ($\dot{M}\sim$10$^{-5}$M$_{\odot}$/yr and 
v$_{\infty}$$\sim$1200 km/s) the X-ray luminosity created by colliding winds 
is of the order of 10$^{32-33}$erg/s, for D$\gtrsim$10-500AU. In this context, it is plausible  
that the enhanced X-ray activity might mark the periastron passage of a system with an
eccentric orbit.
Recently, \citet[][]{2010ApJ...710..706M} reported X-ray emission from massive stars in the 
GC region. Among the sources studied in their sample, four are located in the AC, 
two of them are inside the XMM PSF and are good candidates for the location of the flaring event.
Another example of variability involving early-type stars can be found in the binary system 
$\eta$ Carinae; this has been measured to be variable with an increase in the X-ray
luminosity of about 80\%. In this supermassive colliding wind
binary, the X-ray luminosity reaches 10$^{34}$ erg/s and the flare
timescales range between a few days and about 100 days \citep[][]{2009ApJ...707..693M}. 
A likely explanation for this variability is the interaction of clumpy winds; 
this process has been observed to produced an X-ray luminosity 1.8 times the normal level, 
with a spread in the duration of the high state activity between $\sim$3 and 50 days, characteristics which might equally well describe the AC flare.

Here we propose that the colliding wind binary hypothesis
provides the most plausible explanation for the AC flare. On the basis of the lightcurve,
the upper limit for the duration of the measured flare 
is about four years. Assuming a wind velocity of 1200 km/s, the linear size 
of the region intersected by a propagating density enhancement over a 4-year period 
is about 0.005 pc, whereas the mean 
distance between stars in a dense environment such as the AC core can be 
estimated to be $\sim$ 0.1 pc.
This argument lends further support to the stellar origin of this X-ray activity.
Presumably, such interactions  are quite probable in a
dense cluster like the AC, especially in the southern component, where
the highest concentration of massive colliding wind binaries has been
reported \citep[][]{2002ApJ...570..665Y}.

In the future it will be of great importance to better characterise the temporal variability
exhibited by massive star clusters such as the AC, both from the perspective of the underlying
processes but also in the context of the contribution of such events to the high energy budget
of the whole GC region.

%\bibliographystyle{aa}
%\bibliography{paper.bib}

% =============================================================================
%\appendix
% =============================================================================

\end{document}